\documentclass[10pt,conference]{IEEEtran}
\usepackage{url}
\usepackage{setspace}
\usepackage{etoolbox} 
\AtBeginDocument{%
  }
\usepackage{multirow}
\usepackage{algorithm} 
\usepackage{algorithmicx}
\usepackage{algpseudocode}
\usepackage{multirow}
\usepackage {witharrows}
\usepackage{amssymb}
\usepackage{CJKutf8}
\usepackage[framemethod=TikZ]{mdframed}
\usepackage{amsmath}
\usepackage{booktabs}
\usepackage{xcolor} 
\usepackage{hyperref}

\pdfoutput=1
\definecolor{verylightgray}{RGB}{245,245,245}

\begin{document}

\title{Human-Like Code Quality Evaluation through LLM-based Recursive Semantic Comprehension}

\author{
\IEEEauthorblockN{Fangzhou Xu\IEEEauthorrefmark{1}, Sai Zhang\IEEEauthorrefmark{2}, Zhenchang Xing\IEEEauthorrefmark{3}, Xiaowang Zhang\IEEEauthorrefmark{4}, Yahong Han\IEEEauthorrefmark{5}, Zhiyong Feng\IEEEauthorrefmark{6}}
\IEEEauthorblockA{\IEEEauthorrefmark{1}School of Computer Science and Technology, Tianjin University, Tianjin, China\\
Email: xu\_fangzhou@tju.edu.cn}
\IEEEauthorblockA{\IEEEauthorrefmark{2}School of Computer Science and Technology, Tianjin University, Tianjin, China\\
Email: zhang\_sai@tju.edu.cn}
\IEEEauthorblockA{\IEEEauthorrefmark{3}CSIRO’s Data61, CSIRO, Australia\\
Email: zhenchang.xing@data61.csiro.au}
\IEEEauthorblockA{\IEEEauthorrefmark{4}School of Computer Science and Technology, Tianjin University, Tianjin, China\\
Email: xiaowangzhang@tju.edu.cn}
\IEEEauthorblockA{\IEEEauthorrefmark{5}School of Computer Science and Technology, Tianjin University, Tianjin, China\\
Email: yahong@tju.edu.cn}
\IEEEauthorblockA{\IEEEauthorrefmark{6}School of Computer Science and Technology, Tianjin University, Tianjin, China\\
Email: zyfeng@tju.edu.cn}
}


\maketitle
\begin{abstract}

Code quality evaluation involves scoring generated code quality based on a reference code for a specific problem statement. Currently, there are two main forms of evaluating code quality: match-based evaluation and execution-based evaluation. The former requires the collection of a large number of test cases, making a huge cost. The latter relies on superficial code matching as an evaluation metric, which fails to accurately capture code semantics. Moreover, extensive research has demonstrated that match-based evaluations do not truly reflect code quality. With the development of large language models (LLMs) in recent years, studies have proven the feasibility of using LLMs as evaluators for generative tasks. However, due to issues like hallucinations and uncertainty in LLMs, their correlation with human judgment remains at a lower level, making the direct use of LLMs for code quality evaluation challenging. To address these issues, we propose Human-Like Code Quality Evaluation through LLM-based Recursive Semantic Comprehension (HuCoSC). We employ a recursive approach to enable LLMs to comprehend portions of code semantics independently each time, obtaining the code semantics through multiple interactions with LLMs. We designed a Semantic Dependency Decoupling Storage to make independent analysis feasible, allowing LLMs to achieve more accurate semantics by breaking down complex problems. Finally, the generated code is scored based on a semantic comparison between the reference code and itself. Experimental results indicate that HuCoSC surpasses existing state-of-the-art methods in terms of correlation with human experts and correlation with code execution.

Index Terms---Code evaluation, Large Language Models, Code Semantic
  
\end{abstract}





\section{Introduction}
Code quality evaluation involves scoring generated code quality based on a reference code for a specific problem statement. This evaluation has many applications in software engineering, such as providing coding standards for programming competitions~\cite{liu2023judges}, assisting students in programming learning~\cite{skalka2023automatic}, and offering automatic evaluation of generation quality in training generative models~\cite{lu2021codexglue}~\cite{le2022coderl}~\cite{zan2023large}.

Methods of code quality evaluation can be divided into execution-based and match-based methods~\cite{dong2023codescore}. 
Execution-based code quality assessment typically requires a large number of manually written test cases to evaluate code quality based on the outcomes of code execution~\cite{zhou2023codebertscore}, such as Pass@k~\cite{kulal2019spoc} and AvgPassRatio~\cite{hendrycks2021measuring}. The high costs of the execution-based method lead to a higher frequency of match-based code evaluations in practice~\cite{evtikhiev2023out}. Many code generation models in the open-source community rely on token-match-based metrics like BLEU~\cite{papineni2002bleu}~\cite{evtikhiev2023out} to evaluate the quality of generated code. However, code quality evaluation methods based on token matching often fail to accurately reflect the quality of the code. Many scholars~\cite{rouge2004package}~\cite{tran2019does}~\cite{ren2020codebleu}~\cite{popovic2015chrf} have pointed out the lack of correlation between human experts and token-match-based metrics like BLEU in code quality evaluation and have attempted to bridge this gap. However, the methods proposed are all variants of token-match-based approaches, which offer limited improvements.

With the development of deep learning in recent years, the advent of pre-trained language models has made it possible for machines to comprehend code semantics. CodeBertScore~\cite{zhou2023codebertscore} utilizes the pre-trained model Bert~\cite{zhou2023codebertscore} to obtain semantic vectors of reference code and generated code, and the score of the generated code is derived from the similarity of these vectors. However, since Bert encodes semantic vectors based on context, and the similarity of the context does not necessarily represent the similarity of code semantics, this leads to CodeBertScore not performing well in evaluating codes that are semantically identical but implemented differently. Zhuo proposed a code quality evaluation framework~\cite{Zhuo2023large} based on a large language model (LLM), assuming that the LLM has already learned relevant code knowledge and has the ability to assign higher scores to codes of higher quality. However, most of the codes in the CoNaLa~\cite{yin2018learning} subset of the dataset~\cite{evtikhiev2023out} used for evaluation are single-line codes without complex semantics; although the Card2Code Hearthstone~\cite{ling2016latent} subset consists of semantically more complex {``classes,''} these {``classes''} all share the same structure with very little variation. In reality, a substantial portion of code exhibits complex and varied semantics, and obtaining a high-quality reference code for problem-specific code quality evaluation is very difficult and expensive~\cite{zan2023large}. There is often a significant difference between reference codes and generated codes. Extensive empirical evidence indicates that LLM-based evaluators still have a low correlation with human judgment~\cite{zhong2022towards}, and due to issues like hallucinations and uncertainties~\cite{ji2023survey}~\cite{creswell2022faithful}, directly using LLMs to analyze codes with complex semantics is challenging. Therefore, a more effective and reliable framework is needed to utilize LLMs for code assessment.

To better match the semantic differences between codes, we propose a framework called \textbf{Hu}man-Like \textbf{Co}de Quality Evaluation through LLM-based Recursive \textbf{S}emantic \textbf{C}omprehension (\textbf{HuCoSC}). We designed a Semantic Dependency Decoupling Storage to make independent comprehension possible, allowing LLMs to achieve more accurate semantics by breaking down complex problems. Ultimately, the generated code is scored based on a semantic comparison between the reference code and itself.






We evaluate HuCoSC from two aspects: its correlation with human experts and its correlation with code execution. For the former, we selected 200 problems from the CodeNet dataset~\cite{puri2021codenet}, which is known for its complex semantic competitive programming~\cite{nair2020increasing} problems. We invited four experts to score the codes to be evaluated from 0 to 4 points based on the task description and reference code, using these scores as the evaluation benchmark. On this dataset, HuCoSC, using GPT-3.5 Turbo and GPT-4 Turbo as its backbone models, achieved Spearman correlations of 0.769 and 0.853, respectively, surpassing all existing evaluation methods. For the latter, we conducted experiments on the HumanEval~\cite{humaneval} dataset. HuCoSC, utilizing GPT-3.5 Turbo and GPT-4 Turbo as backbone models, reached Spearman correlations of 0.594 and 0.753, respectively, which exceeded all other methods. 


The main contributions of this paper are as follows:

\begin{itemize}
    \item We propose an automatic code evaluation framework, HuCoSC, which employs a recursive approach to enable LLMs to independently comprehend portions of code semantics each time, obtaining the code semantics through multiple interactions with LLMs. The framework then evaluates the generated code by scoring it based on a semantic comparison between the reference and the generated code.

\end{itemize}

\begin{itemize}
    \item Inspired by human comprehension processes, we designed a Semantic Dependency Decoupling Storage to make independent analysis feasible, allowing LLMs to achieve more accurate semantics by breaking down complex problems.
\end{itemize}

\begin{itemize}
    \item We have collected a new validation dataset for code quality evaluation, Code-Pair, featuring codes with complex semantics and diverse forms. 
\end{itemize}

\section{MOTIVATION}
\subsection{High score on code with semantic errors}
\begin{figure}[h]
  \centering
  \includegraphics[width=1\linewidth]{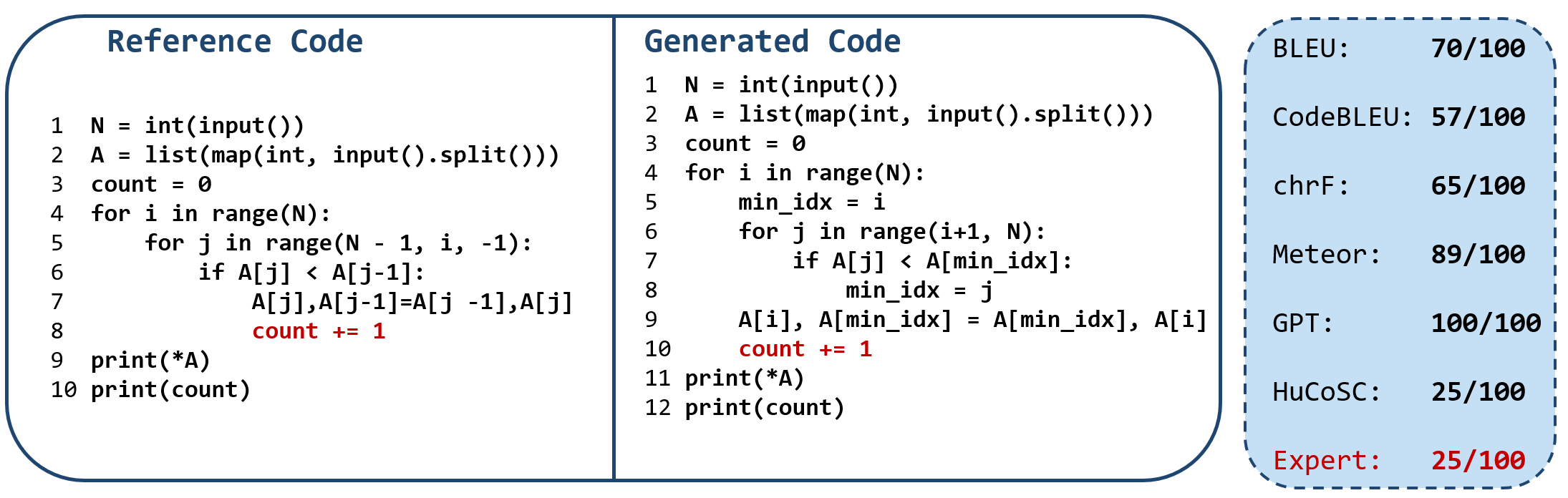}
  \caption{A comparison of scores from different code quality evaluation methods, with the scores of each method converted to a percentage scale.}
  \label{fig_mov1}
\end{figure}
As shown in Fig~\ref{fig_mov1}, the generated code has a logical error in its counting logic, always outputting the array's length. According to the evaluation criteria proposed by Evtikhiev et al.\cite{evtikhiev2023out}, experts rated it 25 points. However, with many variables having the same names, leading to traditional methods achieving higher scores: BLEU 70, chrF 65, and Meteor 89. Since there are slight differences in the loop structures of the two codes, CodeBLEU, which can recognize structural differences in code, scored it 57, which is still much higher than the expert score. Using GPT3.5-Turbo also fails to accurately identify the semantic differences between the two codes, resulting in a full score. However, HuCoSC can correctly identify the semantic error in the generated code and assigns a low score.
\subsection{Low score on code with correct semantic}
\begin{figure}[h]
  \centering
  \includegraphics[width=1\linewidth]{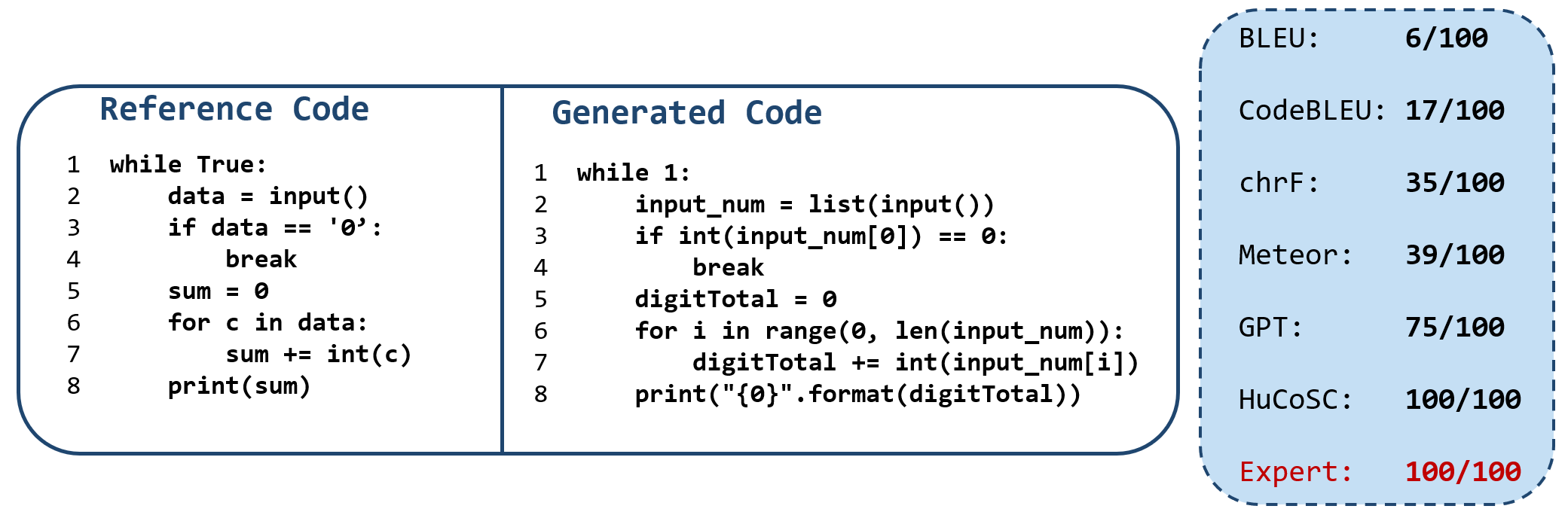}
  \caption{A comparison of scores from different code quality evaluation methods, with the scores of each method converted to a percentage scale}
  \label{fig_mov2}
\end{figure}
As shown in the Fig~\ref{fig_mov2}, both codes perform the function of summing the input array and outputting the result, repeating this operation until the input is 0. The generated code should score 100. However, due to the limited common parts, the scores from traditional methods are very low, with the highest score being only 39 from Meteor. GPT-3.5 Turbo suggests that due to the absence of a loop termination condition, it should be awarded 75 points, but the termination condition was defined in lines 3--4. HuCoSC is able to recognize that the internal semantics of the two codes are the same, awarding 100 points to the generated code.

\section{APPROACH}
\begin{figure}[htbp]
\centerline{\includegraphics[width=1\linewidth]{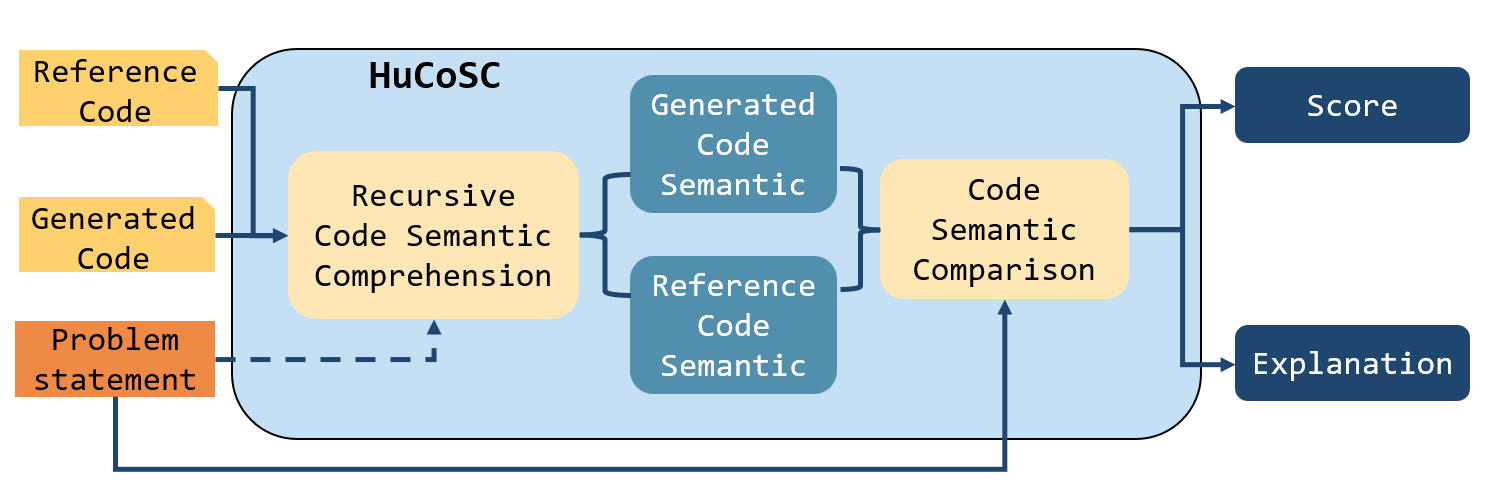}}
\caption{Overall Framework of HuCoSC
}
\label{fig:overall}
\end{figure}
Fig~\ref{fig:overall} illustrates the overall framework of HuCoSC. HuCoSC inputs the generated code, the reference code, and a problem statement. First, the semantic of both codes is obtained through a recursive code semantic comprehension unit. Subsequently, the code semantic comparison unit determines the differences in semantics. Ultimately, the generated code's score and the explanation of the score are derived by analyzing these semantic differences through an LLM, which considers the problem statement. In order to allow the LLM to analyze the code independently without interference from the problem statement, we only introduced the problem statement in a small part of the code semantic comprehension, as shown by the dashed line. In the recursive code semantic comprehension unit, we designed an AI chain~\cite{wei2022chain} mimicking the human process of code comprehension. This approach decomposes the complex task of comprehending code semantics into multiple simple, independent sub-problems. To address the issue of dependencies among sub-problems, we design a Semantic Dependency Decoupling Storage by mimicking human comprehension processes. Each sub-problem interacts with LLMs to progressively ascertain the complete semantics of the code. 

The HuCoSC framework primarily encompasses two key stages: the recursive code semantic comprehension and the code semantic comparison. The recursive code semantic comprehension necessitates the support of both {``Code Recursive Decomposition''} and {``Semantic Dependency Decoupling Storage''} for completion.

\begin{algorithm}
\caption{Recursive Code Semantic Comprehension Pipeline}
\label{alg:Frame-alg}

\begin{algorithmic}[1]
\begin{footnotesize}
\Require $Code$ - the code to be analyzed
\Ensure $Code_{Semantic}$ - semantic descriptions of the code
\Statex \textbf{Data Structure:} 
\Statex \( Storage : \text{Dictionary} \left\{ \text{DependenceName} \mapsto \text{Semantic} \right\} \)

\Function{GetSemantic}{$Code$}
    \State $SubCode_{List} \gets$ \Call{CodeDecomposition}{$Code$}
    \State $SC_{semantics} \gets$ [ ]
    \For{$SC$ in $SubCode_{List}$}
        \State $SC_{Depth} \gets$ \Call{GetDepth}{$SC$}
        \State $DP_{s} \gets$ \Call{GetDependence}{$SC$}
        \State $DP_{semantics} \gets$ \Call{Storage.RetrieveSemantic}{$DP_{s}$}
        \If{$SC_{Depth} < \text{Threshold}$}
            \State $SC_{semantic} \gets$ \textit{LLM}({$SC \oplus DP_{semantic}$})
        \Else
            \State $SSC_{semantic} \gets$ \Call{GetSemantic}{$SC$}
            \State $SC_{semantic} \gets$ \textit{LLM}($SC$$\oplus$$DP_{semantics}$$\oplus$$SSC_{semantic}$)
            \State \ $SC_{semantics}$.append($SC_{semantic}$)
        \EndIf
        \For{$DP$ in $DP_{semantics}$}
            \State $DP_{new} \gets$ \Call{UpdateSemantic}{$DP$, $SC_{semantic}$}
            \State \Call{Storage.UpdateSemantic}{$DP$, $DP_{new}$}
        \EndFor
    \EndFor
    \State $Code_{semantic} \gets$ \Call{SemanticSummarization}{$SC_{semantics}$}
    \State \Return $Code_{semantic}$
\EndFunction
\end{footnotesize}
\end{algorithmic}
\end{algorithm}

\subsection{Code Recursive Decomposition}
Directly analyzing codes' semantic using LLMs is challenging due to issues like errors, hallucinations, and uncertainties~\cite{ji2023survey}~\cite{creswell2022faithful}. Inaccurate comprehension of code semantic leads to an inability to provide precise semantic differences, ultimately resulting in inaccurate code scoring. To mitigate the illusion phenomenon in LLMs and obtain more accurate code semantic, we designed a Code Recursive Decomposition unit to break down the code into several sub-codes for individual analysis by LLMs. There is already considerable work on gradually obtaining code semantics through code decomposition. A popular method involves splitting the Abstract Syntax Tree (AST) to decompose code~\cite{hu2023fine}~\cite{shi2023cocoast}~\cite{choi2023blocsum}. In the AST, the subtree under each {``type name of nodes''} represents a sub-function in the code. However, ASTs are often large and deep~\cite{shi2023cocoast}~\cite{hu2023fine}, and it is impractical to decompose the code at every type of nodes, as such fine-grained decomposition results would cause difficulties in subsequent code semantic summarization. Following the approach by Hu etc. al.~\cite{hu2023fine}, we considered eight types of nodes of AST as our predefined nodes: {``\textbf{For}''}, {``\textbf{While}''}, {``\textbf{Assign}''}, {``\textbf{If}''}, {``\textbf{ClassDef}''}, {``\textbf{FunctionDef}''}, {``\textbf{Switch}''}, and {``\textbf{Call}''}. We perform a depth-first traversal of the code's AST, extracting the {``subtrees''} under these predefined nodes as sub-codes. In this way, we decompose the code into several sub-codes using predefined nodes.

\begin{figure}[h]
  \centering
  \includegraphics[width=\linewidth]{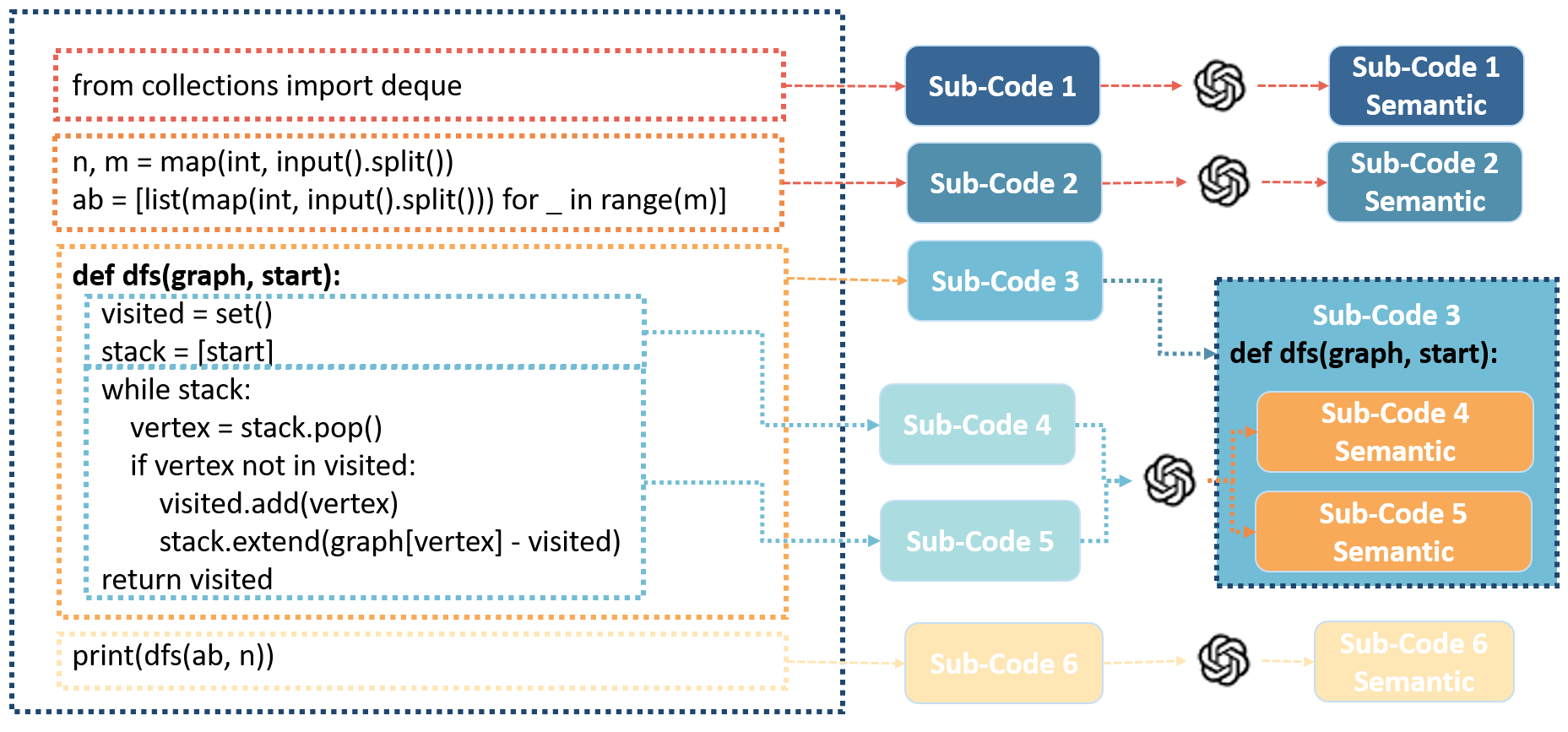}
  \caption{Demonstrates how code is recursively decomposed into sub-codes.}

\label{fig_Code_decom}
\end{figure}

However, all the decomposition methods mentioned previously adopted a single-step decomposition approach, which often does not yield sub-codes of the desired size through the aforementioned one-step decomposition. As illustrated by {``Sub-Code 3''} in Fig~\ref{fig_Code_decom}, Some sub-codes may contain multiple layers of nesting or a large number of lines. Therefore, a further decomposition process is needed to ensure that the sub-codes are optimal in size and complexity for effective comprehension by LLMs.
To quantify the complexity of sub-code, we use {``code depth''} to measure the complexity of sub-codes. {``code depth''} is defined as the nesting level of predefined nodes. As shown in line 2 of Algorithm~\ref{alg:Frame-alg}, upon entering the \Call{GetSemantic}{} function, the code is first decomposed into several sub-codes. Subsequently, if the $SC_{Depth}$ of a decomposed sub-code ($SC$) still exceeds this threshold, as shown in lines 8--11 of Algorithm, the recursive \Call{GetSemantic}{} function continues to perform recursive code decomposition. This process is repeated until the $SC_{Depth}$ of the decomposed $SC$ falls below the threshold. 

As illustrated in Fig~\ref{fig_Code_decom}, {``Sub-Code 3''} contains a nesting of 3 layers of predefined nodes. Assuming a threshold value set to 3, it requires further decomposition. Therefore, the Code Recursive Decomposition unit is called recursively to extract the sub-codes {``Sub-Code4''} and {``Sub-Code5''} under {``Sub-Code 3''}. All code depths are now below the threshold, marking the end of the code recursive decomposition.



\subsection{Semantic Dependency Decoupling Storage}
\begin{figure*}[h]
  \centering
  \includegraphics[width=1\textwidth]{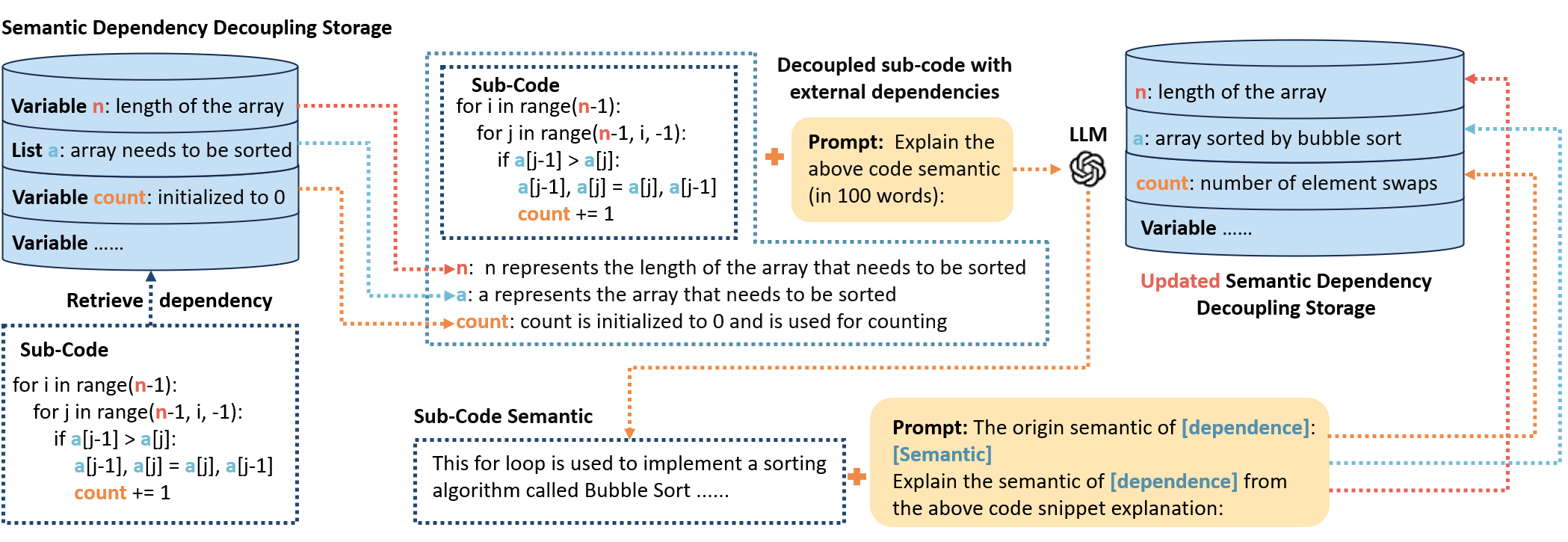}
  \caption{Demonstrates how the Semantic Dependency Decoupling Storage eliminates external dependencies, as well as the process of updating its internal semantic descriptions.}

\label{fig_Code_depend}
\end{figure*}

After decomposing the code into several sub-code, it is not feasible to analyze them individually, as most code segments are interrelated through references and dependencies. Analyzing them in isolation could lead to missing external references, such as variables and function definitions. As shown in Fig~\ref{fig_Code_depend}, direct analysis without knowing the semantics of external variables such as {``n''}, {``a''}, and {``count''} can lead to ambiguity. To address this, we simulate the human approach to comprehension code. When humans comprehend code, they temporarily store the semantics of previously analyzed fragments in their memory. Upon encountering a reference to a previously analyzed segment in another code segment, humans do not reanalyze the previous segment. Instead, they use the semantics stored in their memory to aid in comprehension the current segment's semantics. This use of memory enables humans to analyze individual code fragments without being hindered by interdependencies.



We designed a Semantic Dependency Decoupling Storage that stores textual descriptions of semantics during the comprehension process, which may be required for subsequent code semantic comprehension, such as the semantics of variables, functions, and classes. This unit decouples the dependencies between different sub-codes. As shown in Algorithm~\ref{alg:Frame-alg}, the data structure of the Semantic Dependency Decoupling Storage is a dictionary with dependency names as indexes and semantic descriptions as values. As shown in line 6, we first employ a \Call{GetDependence}{} function to extract its external dependencies ($DP_s$) before analyzing the $SC$ fragment. If the code is semantically correct, each external dependency should have been previously analyzed and exist in the Semantic Dependency Decoupling Storage unit. We simply need to search and retrieve the semantics associated with each external dependency variable. The semantic descriptions ($DP_{semantics}$) of these external dependencies, retrieved from the Semantic Dependency Decoupling Storage unit, is then combined with the $SC$ and inputted into the \textit{LLM} for analysis. This approach eliminates the interdependencies among the sub-codes, allowing for a more accurate and isolated analysis of each $SC$.

As shown in Fig~\ref{fig_Code_depend}, a search is conducted in the Semantic Dependency Decoupling Storage to retrieve related semantic descriptions. These descriptions are concatenated with the original sub-code and then, along with a pre-designed prompt template, are inputted into the LLM to obtain the semantic description of the sub-code.

\subsection{Recursive Code semantic Comprehension}
With Code Recursive Decomposition and Semantic Dependency Decoupling Storage, we can decompose code into sub-codes that are more comprehensible to LLMs and resolve interdependencies among them. Next, we can employ LLMs to perform semantical analysis on each sub-code. Recursive code semantic comprehension can be divided into three main steps: {``Code Semantic Comprehension,''} {``Code Semantic Storage Updating,''} and {``Code Semantic Summarization.''}

\subsubsection{Code Semantic comprehension}

 Having resolved the issue of interdependencies between sub-codes, we are now able to analyze the semantic of each sub-code individually. as shown in lines 8--13 of the Algorithm~\ref{alg:Frame-alg}, if further decomposition is not required, the semantic description of the sub-code ($SC_{semantic}$) can be obtained by inputting the $SC$ combined with the $DP_{semantics}$ to LLM. If further decomposition is required, recursively invoke the \textit{GetSemantic } function to get {``Sub-Sub-Code''} (\textit{SSC})'s semantic description ($SSC_{semantic}$). Merge the $SSC_{semantic}$ with the \textit{SC} and, along with the $DP_{semantics}$, input them into LLM to obtain the semantic description of the sub-code ($SC_{semantic}$).

As shown in Fig~\ref{fig_Code_decom}, assuming the threshold is set to 3, {``Sub-Code''} 1, 2 and 6 do not require further decomposition and can directly prompt\footnotemark[1] the LLM to generate their semantic descriptions after eliminating external dependencies. The depth of {``Sub-Code 3''} exceeds the threshold, necessitating further decomposition into {``Sub-Code''} 4 and 5 according to the previously defined decomposition rules. After eliminating dependencies and prompting the LLM, we obtain the semantic descriptions for {``Sub-Code''} 3 and 4. By concatenating the semantic descriptions with the recursively upper-level code, we achieve the internally recursed {``Sub-Code 3''}, and subsequently, by prompting the LLM again, we obtain the overall semantics of {``Sub-Code 3''}.


An important detail to note is that we only utilize the problem statement to aid the comprehension of LLMs in sub-code with input statements. In other sub-codes, we do not allow LLM to rely on the problem statement for code semantic comprehension. This is because we found that adding additional problem statement information exacerbates the illusion phenomenon when LLMs comprehend code. LLMs analyze the code according to the problem statement. In many cases, code that originally does not align semantically with the problem statement is analyzed as being consistent with it. Especially for models with relatively lower performance (such as GPT-3.5 Turbo~\cite{GPT-3.5Turbo}), adding problem statement does not assist but rather interferes with the LLMs' ability to comprehend and analyze. We will conduct a detailed analysis of this phenomenon in the experimental results. Therefore, we only introduce the problem statement in the code input part to provide an initial context for the code, such as: {``n represents the length of the input array,''} and {``a represents the input array.''} In this way, we incorporate the background information of the problem statement to assist the LLMs in the code semantic comprehension unit while minimizing interference with their independent code analysis.
\subsubsection{Semantic Dependency Decoupling Storage Updating}
The semantic descriptions in the \textit{Semantic Dependency Decoupling Storage} are not static after being defined. For example, during the execution of a program, variables are continuously called and updated, and their semantics change with the running of the code. Therefore, it is necessary to maintain the updates of the \textit{Semantic Dependency Decoupling Storage} unit constantly. As shown in lines 16-17 of Algorithm~\ref{alg:Frame-alg}, each time the semantic description of a sub-code is obtained, the LLM is prompted\footnotemark[1] to obtain the updated semantic descriptions ($DP_{new}$) of each external variable using the description. The updated semantic descriptions are then re-stored in the \textit{Semantic Dependency Decoupling Storage} using the \Call{UpdateSemantic}{} function.

As shown in Fig~\ref{fig_Code_depend}, after obtaining the semantic descriptions of the sub-codes, a well-designed prompt template is used to prompt the LLM to update the semantics of each external dependency. For example, the semantic description of list {``a''} is updated from {``the array that needs to be sorted''} to {``the array after being sorted by bubble sort,''} the semantic description of the variable {``count''} is updated from {``count is initialized to 0 and is used for counting''} to {``stores the number of element swaps.''} The semantic of variable {``n''} remain unchanged, so it is kept as is.

\subsubsection{Code Semantic Summarization}
After recursive decomposition and semantic comprehension, we have obtained the semantics of each sub-codes, As shown in line 20 of Algorithm~\ref{alg:Frame-alg}, after obtaining the semantics of each sub-code, we use the function \Call{SemanticSummarization}{} to obtain the overall semantics ($Code_{semantic}$) of the code. As shown in the Fig~\ref{fig_Code_decom}, sub-codes semantic descriptions like {``Sub-Code 1 Semantic''}, {``Sub-Code 2 Semantic''}, etc. We concatenate these {``Sub-Code semantics''} and use a designed prompt template\footnotemark[1] to enable the LLM to deduce the overall semantics of the code. Similarly, to prevent problem statement from interfering with the summarization of semantics, we did not introduce problem statement to assist LLMs comprehension.
\subsection{Code Semantic Comparison}
After obtaining the semantic description of the reference code and the generated code, we designed a Code Semantic Comparison unit. This unit enables the LLM to differentiate the semantic differences between the two codes, verifying whether the generated code can achieve functions as efficiently as the correct code by using a prompt template\footnotemark[1]. Subsequently, based on user-defined evaluation criteria and considering the semantic differences between codes in conjunction with the problem statement, the LLM is prompted to score the generated code. 
We designed the evaluation criteria to be modifiable by users, as different evaluation tasks have different focuses. Some tasks emphasize the correctness of the code's logic, while others prioritize whether the code executes correctly. Moreover, we leverage the characteristics of LLMs to not only generate scores but also explain the reasoning behind these scores, which aids in the application of our method in areas like programming education and programming assistance. We utilize Few-Shot prompts\footnotemark[1] to standardize the output of LLMs, enabling the use of regular expressions to separate scores from explanations.

\section{EXPERIMENTAL DESIGN}

\subsection{Research Questions}

In the experimental section, we aim to address the following research questions (RQs):
\begin{itemize}
    \item \textbf{RQ1}: How does HuCoSC perform compared to other code quality evaluation methods?
\end{itemize}
\begin{itemize}
    \item \textbf{RQ2}: How does recursive code semantic comprehension unit perform?
\end{itemize}
 \begin{itemize}
    \item \textbf{RQ3}: What impact does the problem statement have on code comprehension?
\end{itemize}
 \begin{itemize}
    \item \textbf{RQ4}: What is the quality of the explanations generated by HuCoSC?
\end{itemize}

\subsection{Data Preparation}
\subsubsection{Code-Pair}
We collected 85 Competitive Programming (CP) tasks from the CodeNet~\cite{puri2021codenet} dataset as evaluation examples, acting as a validation set for correlation with human experts. CP tasks, characterized by their strong semantic properties, require participants to offer programming solutions based on problem statements~\cite{nair2020increasing}, focusing primarily on algorithms, data structures, and time complexity~\cite{das2022exploring}. To simulate real-world code evaluation scenarios as closely as possible, we also employed GPT-3.5 Turbo~\cite{GPT-3.5Turbo} and the Mistral-7B~\cite{puri2021codenet} model to generate code based on CP problem statements, further enriching our dataset. The code generated by GPT-3.5 Turbo simulates higher-quality code as encountered in real-world scenarios, whereas Mistral-7B represents code of moderate to lower quality. This approach provides a diverse and realistic setting for assessing and testing the robustness of our method. We collected a validation dataset, Code-Pair, consisting of 428 reference--generated code pairs. Within this dataset, the reference codes were selected from CodeNet, with 193 pairs of generated code were generated using GPT-3.5 Turbo, and 235 pairs were generated using the Mistral-7B model.

For each code pair, we had four experts, each with over four years of Python programming experience, score the quality of the generated code. In this process, each expert can utilize the test cases included in the CodeNet problem statements to test the code when scoring, thereby assisting the experts in their evaluation. The code pairs will be evaluated by experts in a random order. The experts will give the evaluation code 0-4 points based on the problem statement, correct code and following evaluation criteria:

    \textbf{0.Functionality cannot be achieved} - The generated code is completely irrelevant to the problem statement.
    
    \textbf{1.Semantic Error} - The generated code is relevant to the problem statement, but has serious semantic problems.
    
    \textbf{2.Partial semantic functionality realized} - Generated code may be able to perform some of the functionality, but does not cover all functional situations or has incomplete inputs and outputs.
    
    \textbf{3.Functionality is realized but are insufficient} -Implementation is less efficient than correct code, with higher time or space complexity.
    
    \textbf{4.Functionality is fully implemented} - the generated code has the same functionality and semantic as the correct code, and has the same execution efficiency, with complete inputs and outputs.

\begin{figure}[h]
  \centering
  \includegraphics[width=\linewidth]{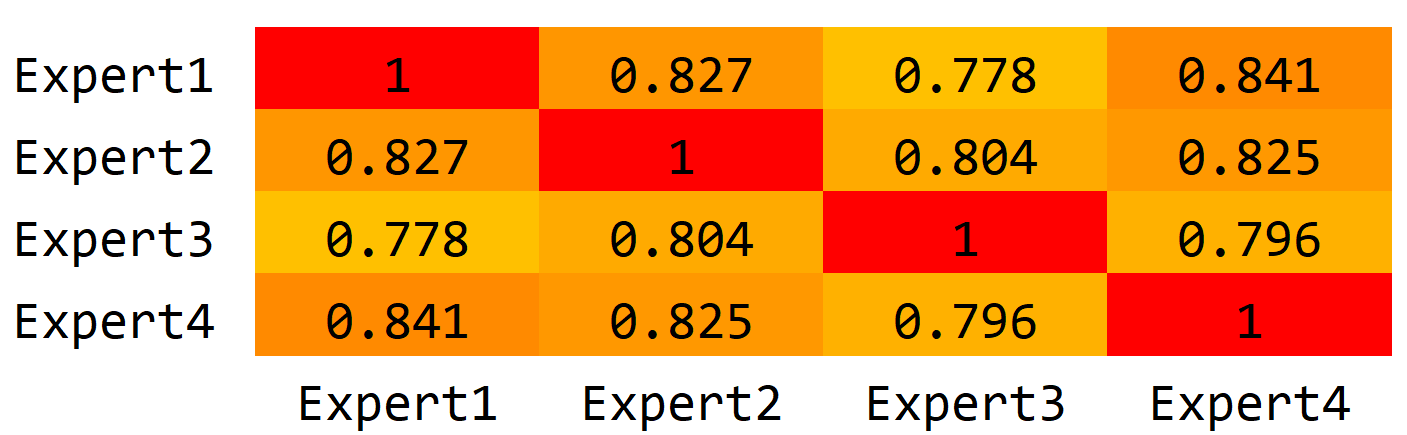}
  \caption{The Kappa coefficients between the four experts.}

\label{fig_expertCorr}
\end{figure}

We used the Kappa coefficient to measure the consistency among the experts. As shown in ~\ref{fig_expertCorr}, The Kappa coefficients between the four experts are all around 80\%, which is a high degree of consistency.

\subsubsection{HumanEval}
The HumanEval~\cite{humaneval} dataset is used as the validation set for correlation with code execution. This dataset includes problem statements, input-output examples, and manually written example codes. Cassano et al.~\cite{10103177} ran test cases and provided the corresponding functional correctness of the programs. Due to the extensive size of the HumanEval dataset, we randomly selected 500 cases to serve as our validation set.
\subsection{Implementation Details}
We use OpenAI's GPT series as our LLMs, including GPT-3.5 Turbo (gpt-35-turbo-1106~\cite{GPT-3.5Turbo}), GPT-4 (gpt-4~\cite{GPT-4}), and GPT-4 Turbo (gpt-4-1106-preview~\cite{GPT-4Turbo}). The temperature parameter controlling the randomness of the LLMs' output is uniformly set to 0.2 to obtain more deterministic answers. The threshold of recursive decomposition is set to 3.
To prevent LLMs from comprehending code semantics based on code comments, we uniformly remove the comments of each code.
\subsection{Baselines}

We compared HuCoSC with existing code quality evaluation methods, encompassing traditional n-gram match-based methods and those utilizing LLMs. Given the limited use of LLMs in code evaluation to date, to ensure fairness in assessment, we also included well-recognized, effective prompt-based methods for comparison:

\subsubsection{Match-based method}
Match-based method include \textbf{BLEU}~\cite{ren2020codebleu}, \textbf{Rouge}~\cite{rouge2004package}, \textbf{METEOR}~\cite{banerjee2005meteor},  \textbf{chrF}~\cite{popovic2015chrf}, \textbf{CodeBleu}~\cite{ren2020codebleu}.

\subsubsection{LLM-based method}

LLM-based method include 

\textbf{CodeBertScore}~\cite{zhou2023codebertscore}:  Based on a BERT~\cite{devlin2018bert} variant, CodeBERTScore is a specialized evaluation method for code generation. It leverages deep learning models to comprehend the semantic and contextual information of code. 

\textbf{Zhuo's Method}~\cite{Zhuo2023large}: Two approaches for code quality evaluation using LLMs: With Reference Code\textbf{(w/R.)}: Inputs include problem statement, generated code, and reference code. LLMs compare the generated code against the reference code based on prompts\footnotemark[1]. Without Reference Code\textbf{(w/o R.)}: Only the problem statement and the generated code are used. LLMs score the code without reference code comparison. The prompt template informs the LLM that it needs to complete a code evaluation task and then add human-defined analytical steps at the end to assist the LLM in gradually analyzing the code and scoring. 

\textbf{LLMs with Chain of Thought (CoT)}~\cite{kojima2022large}: Prompts the LLM to explain the reasoning or steps for a problem before coding. Adding {``Let's think step by step and figure it out''} at the end of prompt\footnotemark[1].

\textbf{LLMs with Few-Shot Learning}~\cite{brown2020language}:  Provides the LLM with a few examples before asking it to perform a task, evaluating the model's ability to adapt its comprehension based on these examples. We use five code evaluation cases as a few-shot sample\footnotemark[1], and each of these five code evaluation examples contains one scenario ranging from 0 to 4 points. 

\textbf{Simplified HuCoSC}: 
We offer a simplified version of HuCoSC, as shown in Fig~\ref{fig:overall}. The simplified HuCoSC replaces the recursive code semantic comprehension unit by directly obtaining the semantics of the code by prompting\footnotemark[1] the LLM.

\subsection{Evaluation Metrics}
\subsubsection{Correlation Metrics}
We evaluate the effectiveness of HuCoSC and each method through the correlation with human experts. We use two metrics: Kendall-Tau~\cite{kendall1938new} and Pearson~\cite{cohen2009pearson} to measure the correlation with human experts of each method. The correlation index value ranges between -1 and 1. A correlation value approaching 1 indicates a strong positive correlation between the two variables. Conversely, a correlation value nearing -1 denotes a strong negative correlation. A value close to 0 suggests no direct relationship between the two variables.
\subsubsection{Scale for human ratings}
In our experiments, we need to subjectively evaluate the quality of intermediate results and the explanations of scoring. We used four python experts to evaluate the results according to a 5--point Likert scale~\cite{joshi2015likert}. Each expert possesses over four years of experience in writing Python programs.
\section{RESULTS AND ANALYSIS}
\subsection{\textbf{RQ1}: The performance of HuCoSC}

\begin{table}

\setlength{\abovecaptionskip}{10pt}
\caption{Kendall-Tau ($\tau$), Pearson ($r_s$) correlations. \textbf{(w/o P.)}:without utilizing Problem statement information;\textbf{(w/o R.)}: without reference code; \textbf{(w/ R.)}: with reference code. The best performance is\textbf{ bold}.}
\label{tab:campare}
\centering
\begin{tabular}{|l|c|c|c|c|}
\hline
Method & \multicolumn{2}{c|}{Code-Pair} & \multicolumn{2}{c|}{HumanEval} \\
\cline{2-5}
&$r_s$ &$\tau$ & $r_s$ &$\tau$ \\
\hline
BLEU   & .226 & .171 &.345 &.282 \\ 
Rouge-1 & .260 & .199 &.383 &.313 \\ 
    Rouge-2 & .193& .147 &.313 &.264 \\
    Rouge-L & .284& .217 &.374 &.306\\
    METEOR & .267& .199 &.323 &.264\\
    chrF & .326& .247 &.330 &.269\\
    CodeBleu & .317& .239 &.295 &.241\\
    CodeBertScore & .315& .237 &.430 &.352\\
    GPT-3.5 Turbo (Zhuo w/o R.) & .529 & .466 &.091 &.090\\
    GPT-3.5 Turbo (Zhuo w/R.) & .572 & .492 &.106 &.105\\
    GPT-3.5 Turbo (CoT) &  .589 & .508 &.142 &.136\\
    GPT-3.5 Turbo (few-shot) & .560 & .464 &.186 &.184\\
    Simplified HuCoSC (GPT-3.5 Turbo) & .626 & .512 &.512 &.470\\
    HuCoSC (GPT-3.5 Turbo) & \textbf{.769} & \textbf{.662} &\textbf{.\textbf{594}} &\textbf{.553}\\
    HuCoSC (GPT-3.5 Turbo w/o P.) & .739 & .646 &.536 &.502\\
    GPT-4 Turbo (CoT) & .678 & .580 &.306 &.285\\
    GPT-4 Turbo (Zhuo w/o R.) & .649 & .557 &.294 &286\\
    GPT-4 Turbo (Zhuo w/R.) & .727 & .622 &.317 &.308\\
    GPT-4 Turbo (few-shot) & .746 & .640 &.289 &.280\\
    GPT-4 Turbo (CoT) & .747 & .640 & .306&.285\\
    Simplified HuCoSC (GPT-4 Turbo) & .761 & .667 &.559 &.527\\
    HuCoSC (GPT-4 Turbo) & \textbf{.853} & \textbf{.780} &\textbf{.753} &\textbf{.718}\\
    HuCoSC (GPT-4 Turbo w/o P.) & .841 & .761 &.656 &.596\\
    Expert & 0.97&0.91  & -&-\\
\hline
\end{tabular}
\end{table}

\subsubsection{Motivation}
We are interested in how HuCoSC compares to other existing methods for evaluating code quality. Our goal is to compare our method with existing code evaluation methods to see whether it offers a superior correlation with human expert judgments and code execution.

\subsubsection{Methodology}
We employ a range of existing techniques for comparison with our approach, including n-gram matching, code structure matching, deep learning-based semantic matching of code, and semantic matching based on large-scale models. To ensure the fairness of our experiments, we also tested the effectiveness of the currently recognized prompt-based approaches on this task. The Code-Pair dataset reflects the correlation of the methods with human expert judgments, and the HumanEval dataset reflects the correlation with code execution.
For prompt-based methods, we utilized different evaluation criteria on two datasets and designed corresponding prompt templates for each method to achieve their best performance.
\subsubsection{Result Analysis}
As shown in Table~\ref{tab:campare}, methods based on n-gram matching (such as Bleu, Meteor, Rouge), code structure matching (CodeBleu), and deep learning-based code semantic matching (CodeBertScore) all exhibit low levels of correlation on the Code-Pair and HumanEval datasets. This indicates that superficial matching of code cannot represent the actual semantics or execution results of the code. Many codes that perform the same function can have different implementation methods or writing habits, leading to the possibility that superficial similarity does not necessarily reflect semantic or execution similarity.

The use of LLMs has shown significant improvement over other methods. On the Code-Pair dataset, the Pearson correlation with human experts for methods using CoT and Few-Shot prompts reached 0.589 and 0.560, using GPT-3.5 Turbo respectively, indicating that LLMs can better capture code semantics. Using GPT-3.5 as the backbone model, Simplified HuCoSC and HuCoSC's Pearson correlation with human experts reached 0.626 and 0.769 respectively, significantly, surpassing all existing methods using GPT-3.5 and approaching the performance of methods using GPT-4 Turbo. Additionally, to achieve a better correlation with human experts, we conducted batch experiments using HuCoSC based on GPT-4 Turbo, achieving a correlation of 0.853 with human experts, surpassing all existing benchmarks.

However, on the HumanEval dataset, the performance of LLMs did not meet expectations. The Pearson correlation coefficients using CoT and few-shot prompts were only 0.142 and 0.166, respectively, even lower than some traditional match-based methods. This indicates that due to the limitations of the LLMs, directly comprehending the execution outcomes from code text is quite challenging, resulting in poor performance. Nonetheless, HuCoSC, utilizing GPT-3.5 Turbo and GPT-4 Turbo, achieved Pearson correlation coefficients of 0.594 and 0.753, respectively. Although there was a decline compared to the results on the Code-Pair dataset, it still significantly outperforms existing methods. This is because HuCoSC can gradually comprehend and accurately capture the semantics of the code and then use these semantics to determine whether the code can correctly execute the problem without having to bridge the gap from code text to code execution directly.
HuCoSC achieved a high correlation on both the Code-Pair and HumanEval datasets, indicating that simply modifying the evaluation criteria can adapt the framework to different code evaluation tasks.

There are some tasks with no problem statement information available but requiring code quality evaluation, such as code translation~\cite{roziere2020unsupervised} and unit test generation~\cite{lukasczyk2023empirical}. When HuCoSC employed GPT-3.5 Turbo and GPT-4 Turbo without integrating problem statements, there was a decrease in the Pearson correlation by 0.03 and 0.12 for the Code-Pair dataset, and by 0.058 and 0.097 for the HumanEval dataset, respectively. This suggests that HuCoSC can still perform code evaluation tasks well, even in the absence of no problem statement information.

\vspace{10pt}
\begin{mdframed}[backgroundcolor=verylightgray, linecolor=gray,linewidth = 1.5pt, roundcorner=5pt, leftmargin=0cm, rightmargin=0cm, innertopmargin=5pt, innerbottommargin=5pt]
\textbf{} \\
HuCoSC surpasses all current methods in correlation with human experts and code execution, enabling accurate code evaluation and adaptability to code evaluation tasks with different criteria.
\end{mdframed}

\subsection{\textbf{RQ2}: The quality of Recursive Code Semantic Comprehension Unit}
\begin{figure}[h]
  \centering
  \includegraphics[width=\linewidth]{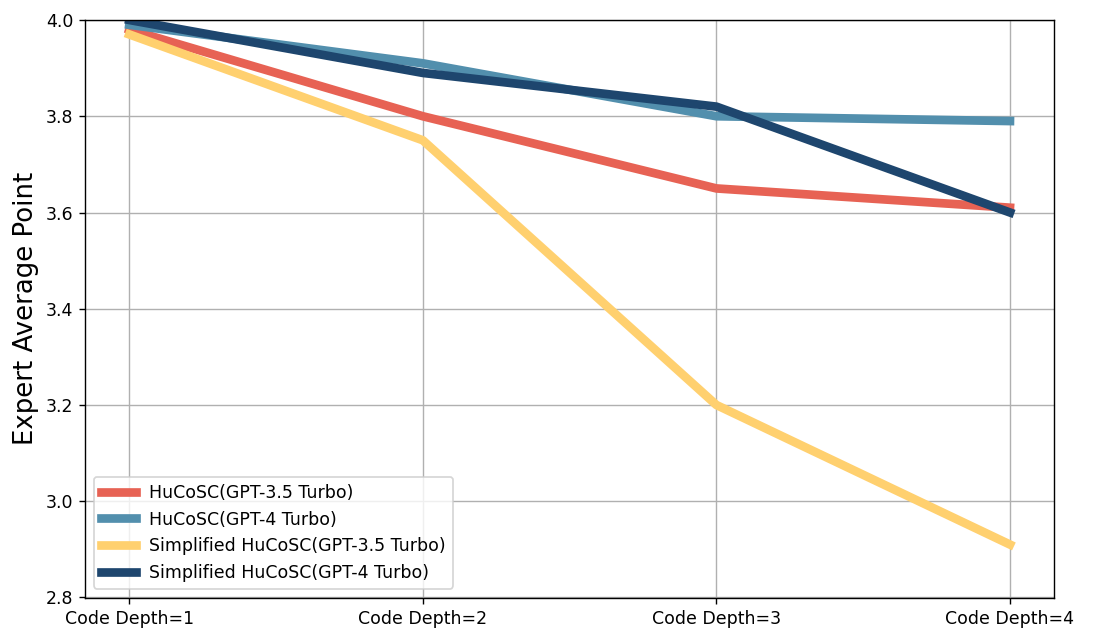}
  \caption{The average expert point for code semantic comprehension at different code depths}
\label{fig_rq2}
\end{figure}

\subsubsection{Motivation}
Since we have designed a chain to simplify complex code to assist LLMs in comprehending code, verifying the intermediate processes' quality is necessary. Our focus here is on whether the semantic functionality of the code, as an intermediate result, is correctly captured, and how HuCoSC improves the effectiveness of code evaluation. 
\subsubsection{Methodology}
We chose 50 code samples each for {``code depths''} of 1, 2, 3, and 4 from the Code-Pair and HumanEval datasets, culminating in 200 codes. 4 human experts were asked to evaluate the quality of the code semantics output by the framework based on a  5-point Likert scale. Through these two experiments, we can explore HuCoSC's ability to comprehend code semantics at different levels of code complexity and the overall performance trends of the framework.

\subsubsection{Result Analysis}
The experimental results, as shown in Fig~\ref{fig_rq2}, indicate that with the increase in code depth, the quality of code semantics and the correlation coefficients of the overall framework both show a general declining trend. Among them, Simplified HuCoSC, due to the absence of a code decomposition step, significantly declines code semantic quality as code complexity increases. The average scores of experts using GPT-3.5 Turbo and GPT-4 Turbo with Simplified HuCoSC drop by 0.7 and 0.19 points, respectively, at a {``code depth''} of 4 compared to HuCoSC. This suggests that the more complex the code semantics, the more difficult it is for LLMs to accurately capture and express them, leading to a more severe hallucination phenomenon. However, HuCoSC, with its recursive decomposition step, divides the code into blocks that are more comprehensible for LLMs, making the impact of code complexity on code semantic quality less significant. This result demonstrates that HuCoSC's approach of recursive decomposition can significantly mitigate the hallucination phenomenon LLMs face when comprehending complex code.


\vspace{10pt}
\begin{mdframed}[backgroundcolor=verylightgray, linecolor=gray,linewidth = 1.5pt, roundcorner=5pt, leftmargin=0cm, rightmargin=0cm, innertopmargin=5pt, innerbottommargin=5pt]
\textbf{} \\
Recursive comprehension of code semantics notably addresses the challenge of accurately capturing semantics as the complexity of code semantics escalates. The improvements brought by HuCoSC are primarily reflected in the evaluation of complex codes.
\end{mdframed}

\subsection{\textbf{RQ3}: The impact of introducing problem statements on HuCoSC}

\subsubsection{Motivation}

As shown in Table~\ref{tab:campare}, we observed a phenomenon where Zhuo's method underperforms when not using reference code compared to some traditional prompt methods. After analyzing the output, we discovered that when scoring without the reference code, LLMs are likely to analyze code semantics based on the problem statement. Without the correct code as a reference, LLMs are prone to misinterpret the semantics of the code, aligning it with the problem statement even if it does not match the original problem semantics. In other words, problem statement can sometimes interfere with LLMs' semantic comprehension of the code, leading to a higher probability of misclassifying the code as semantically correct. We want to analyze how problem statement affect LLM's comprehension of code.
\subsubsection{Methodology}
To investigate the effect of problem statements inclusion on the performance of our framework, we established two control groups: 1) $P_{Full}$: In every step, a {``problem statement''} was introduced as background information to aid the LLM in comprehension code semantics; 2) $P_{Lack}$: No {``problem statement''} was provided at any step, challenging the LLM to grasp code semantics without any background context. $P_{Initial}$: Our approach introduce {``problem statement''} only when the LLM processes sub-codes that include input statements, establishing a foundational context for comprehending the code. Since the code in the HumanEval dataset contains only one function body, we use the Code-Pair dataset, which has a more complex structure, to randomly select 200 pairs of code for experimentation.
\subsubsection{Result Analysis}
\begin{figure}[h]
  \centering
  \includegraphics[width=\linewidth]{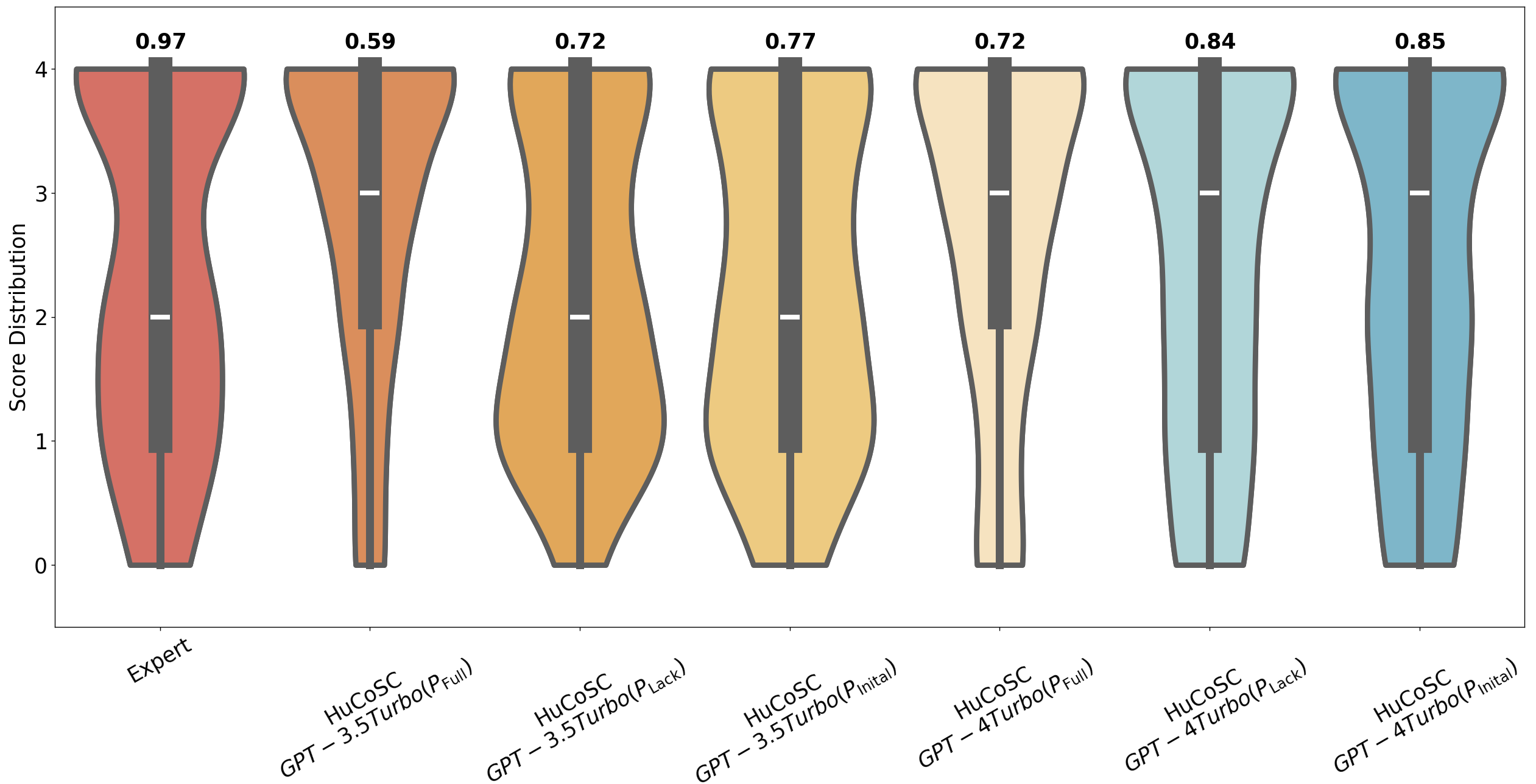}
  \caption{The impact of incorporating problem statements to different extents on the distribution of HuCoSC scoring. The values at the top of each violin plot represent the Pearson correlation coefficient of that group with the experts.}

\label{fig_taskscore}
\end{figure}

As shown in Fig~\ref{fig_taskscore}, Excessive problem statements tend to result in higher scores. After incorporating problem statements into each step of code comprehension with HuCoSC using GPT-3.5 Turbo and GPT-4 Turbo, the correlation with human experts decreased by 0.18 and 0.132 respectively, and the average scores increased by 0.36 and 0.17 respectively. After analyzing the intermediate results, it was found that LLMs tend to analyze sub-codes along the lines of the problem statements, which exacerbates the hallucination phenomenon in LLMs. This leads to some originally incorrect codes being analyzed as conforming to the semantics of the problem statements upon its addition, resulting in higher final scoring. Fig~\ref{fig_taskscore} shows the score distribution violin plots for each experimental group, with the $P_{Full}$ control group having more scores distributed in the higher range, and the proportion of scores between 0-2 significantly reduced.


Comparing $P_{Lack}$ and $P_{Initial}$ (GPT-3.5 Turbo), we observed that providing LLMs with context without any problem statement results in a higher distribution of lower scores. After analyzing the intermediate results, it was found that due to the lack of background information, LLM lacks knowledge of the code's purpose, input variable meanings, etc. This leads to greater randomness in the earlier stages of code semantic analysis and instability in the output of code semantics. In the final semantic comparison, some codes that originally shared the same semantics as the correct codes are judged as having different semantics, resulting in lower scores. However, this phenomenon is almost eliminated when using GPT-4 Turbo, but providing initial semantics is still necessary for the relatively less capable GPT-3.5 Turbo.
\vspace{10pt}
\begin{mdframed}[backgroundcolor=verylightgray, linecolor=gray,linewidth = 1.5pt, roundcorner=5pt, leftmargin=0cm, rightmargin=0cm, innertopmargin=5pt, innerbottommargin=5pt]
\textbf{} \\
Excessive incorporation of problem statements intensifies the occurrence of hallucinations. For less robust models, selectively providing problem statements at specific stages of the analysis to establish an initial context can modestly enhance the LLM's effectiveness in comprehension code semantics.
\end{mdframed}

\subsection{\textbf{RQ4}: Quality of the scoring explanations}
\subsubsection{Motivation}
HuCoSC can not only output the score of the generated code but also offer an explanation of the score according to the evaluation criteria. The explanations of code evaluation have many applications, such as guiding individuals in learning to program~\cite{skalka2023automatic}. In addition to the framework's correlation metrics with human experts, we also focus on the correctness of the generated explanations.
\subsubsection{Methodology}
We conducted a comparative analysis of the evaluation explanations generated by the simplified HuCoSC and the HuCoSC against the CoT prompts. We randomly selected 100 data entries from both the Code-Pair and HumanEval datasets, and the interpretation of their scores was evaluated by experts using a 5--point Likert scale.
\subsubsection{Result Analysis}

\begin{figure}[h]
  \centering
  \includegraphics[width=\linewidth]{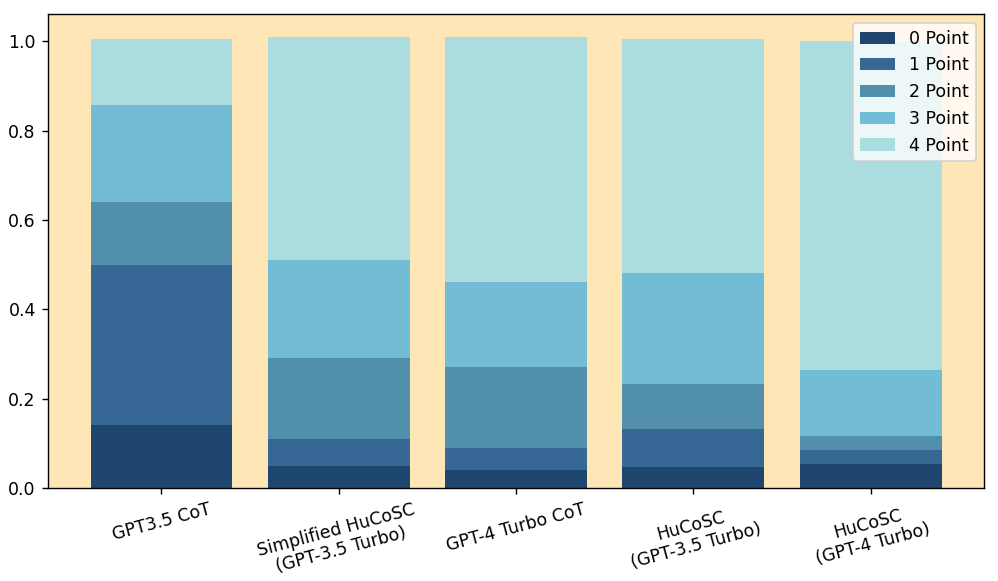}
  \caption{The distribution of expert scores of the scoring explanations generated by different methods}

\label{fig_score_dis}
\end{figure}


The expert score distribution for each method is depicted in the Fig~\ref{fig_score_dis}. The quality of the score explanations obtained from GPT-3.5 using CoT prompts is poor, with nearly half of the explanations scoring below 2 and an average score of only 2.64. However, the situation has significantly improved with the simplified HuCoSC. It can be seen that the proportion of scores at 0 and 1 has noticeably decreased, with effects close to GPT-4 Turbo CoT. Expert ratings are more distributed in the 2--4 range, with average scores reaching 3.02 and 3.11, respectively. This shows that by making LLM focus more on semantic differences, the quality of the explanations can be significantly improved. Using GPT-3.5 as the backbone model, the average expert rating of HuCoSC reached 3.2, and a noticeable decrease in the distribution of lower scores can be observed. This indicates that our method of recursive semantic summarization to obtain more accurate code semantics is effective, as more accurate code semantics can enhance the subsequent judgments and explanations of LLMs. Implementing HuCoSC with GPT-4 Turbo markedly enhances expert consensus on the scoring explanations relative to others. It is evident that all expert ratings, except for those at level 4, have significantly dropped, with a predominant concentration at the 4-point mark, signifying {``strong expert agreement,''}.
\vspace{10pt}

\begin{mdframed}[backgroundcolor=verylightgray, linecolor=gray,linewidth = 1.5pt, roundcorner=5pt, leftmargin=0cm, rightmargin=0cm, innertopmargin=5pt, innerbottommargin=5pt]
\textbf{} \\
While achieving a high correlation with human experts and code execution in its output scores, HuCoSC also ensures the quality of its scoring explanation.
\end{mdframed}

\section{DISCUSSION}
\subsection{Threats to Validity}

HuCoSC currently utilizes the OpenAI API for implementation. However, the API's pricing may not be affordable for everyone, especially during large-scale usage. Due to HuCoSC's good performance on weaker models (like GPT-3.5 Turbo), we plan to validate the effectiveness of HuCoSC using open-source models in the future. This approach will allow the framework to run locally, significantly reducing costs.



\section{RELATED WORK}
\subsection{Token-match-based evaluation}
The earliest methods of code quality evaluation borrowed metrics from machine translation for assessing the quality of generated translations, such as BLEU~\cite{papineni2002bleu}. 
 Although traditional n-gram-based evaluation metrics have been proven to correlate well with human judgment in machine translation~\cite{rouge2004package}~\cite{banerjee2005meteor}, such code evaluation methods often fail to reflect the quality of code accurately in many scenarios.
Tran et al.~\cite{tran2019does} introduced a new code evaluation metric, RUBY, demonstrating that BLEU is ineffective in reflecting the semantic accuracy of translated code and in comparing different SMT-based code migration systems effectively. Shuo et al.~\cite{ren2020codebleu} highlighted the significant differences between code and natural language, suggesting that metrics used for evaluating generation quality in machine translation cannot be directly applied to code generation. They utilized the matching structure of the code, achieving a higher correlation with programmers. However, the similarity of contexts does not necessarily indicate the similarity of code semantics. This leads to the above methods being unable to effectively evaluate semantically identical codes but implemented differently.
\subsection{LLMs-based evaluation}
CodeBERTScore~\cite{zhou2023codebertscore}, by training the masked language model BERT~\cite{devlin2018bert} to encode the context of programs and using the similarity of encoded vectors as scores for generated code. However, since Bert encodes semantic vectors based on context, and the similarity of the context does not necessarily represent the similarity of code semantics, this leads to CodeBertScore not performing well in evaluating codes that are semantically identical but implemented differently

In recent years, the impressive performance of LLMs like GPT~\cite{GPT-3.5Turbo} across various domains has garnered the attention of many scholars for their use in evaluating generative tasks. Chiang et al.~\cite{chiang2023can} highlights that compared to human evaluators, LLMs offer stronger repeatability and independent assessments, and their cost is relatively lower than human experts. Liu et al. developed a framework~\cite{liu2303g} utilizing thought chains to assess the quality of Natural Language Generation outputs using LLMs, surpassing all existing metrics in correlation with human experts. Zhuo introduced a LLM-based framework~\cite{Zhuo2023large} for code evaluation. However, substantial empirical data indicates that evaluators grounded in LLMs maintain relatively low correlation with human judgement~\cite{zhong2022towards}. 

\section{CONCLUSION}
we propose Human-Like Code Quality Evaluation through LLM-based Recursive Semantic Comprehension (HuCoSC). We employ a recursive approach to enable LLMs to comprehend portions of code semantics independently each time, obtaining the code semantics through multiple interactions with LLMs. We designed a Semantic Dependency Decoupling Storage to make independent analysis feasible, allowing LLMs to achieve more accurate semantics by breaking down complex problems. Ultimately, the generated code is scored based on a semantic comparison between the reference code and itself.
The experimental results show that HuCoSC surpasses all existing methods in correlation with human experts and code execution. Furthermore, the experiments validate that recursive decomposition can significantly reduce the hallucination phenomenon experienced by LLMs when comprehending complex code. While achieving a high correlation with human experts and code execution in its output scores, HuCoSC also ensures the quality of its scoring explanation.



\bibliographystyle{IEEEtran}
\bibliography{reference}
\end{document}